\documentclass[apj]{emulateapj}
\usepackage{apjfonts}
\usepackage{amsbsy}

\usepackage{color}
\usepackage{rotating}

\def\om{\Omega_m}

\def\omb{\Omega_b}
\def\s8{\sigma_8}
\def\msol{$M$_\odot}
\def\lcdm{$\Lambda$CDM}

\def\m12{M_{12}}

\def\m12{M_{12}}

\def\rs0{\hat{R}_{\rm sh}^0}
\def\mg2{Mg\,II}

\def\wth{w(\theta)}

\def\mgal{M_{\rm gal}}

\def\mhalo{M_h}

\def\omg{\tilde{\omega}(R)}

\begin{document}

\title{The Correlated Formation Histories of Massive Galaxies and Their Dark Matter Halos}

\author{Jeremy L. Tinker\altaffilmark{1}, 
Matthew  R. George\altaffilmark{2}, 
Alexie Leauthaud\altaffilmark{3}, 
Kevin Bundy\altaffilmark{3},\\
Alexis Finoguenov\altaffilmark{4,5}, 
Richard Massey\altaffilmark{6},
Jason Rhodes\altaffilmark{7,8},
Risa H. Wechsler\altaffilmark{9}
}

\submitted{Submitted to ApJL}
\email{jeremy.tinker@nyu.edu}

\altaffiltext{1}{Center for Cosmology and Particle Physics, Department
  of Physics, New York University}

\altaffiltext{2}{Department of Astronomy, University of California,
  Berkeley, CA 94720, USA}

\altaffiltext{3}{Kavli Institute for the Physics and Mathematics of
  the Universe, Todai Institutes for Advanced Study, the University of Tokyo, Kashiwa, Japan 277-8583 (Kavli IPMU, WPI)}

\altaffiltext{4}{Max Planck Institut f\"{u}r extraterrestrische
 Physik, Giessenbachstrasse, D-85748 Garchingbei M\"{u}nchen,
  Germany}

\altaffiltext{5}{University of Maryland Baltimore County, 1000 Hilltop
  circle, Baltimore, MD 21250, USA}

\altaffiltext{6}{Institute for Computational Cosmology, Durham University, South Road, Durham, DH1 3LE, U.K.}

\altaffiltext{7}{California Institute of Technology, MC 350-17, 1200
 East California Boulevard, Pasadena, CA 91125, USA}

\altaffiltext{8}{Jet Propulsion Laboratory, California Institute of Technology, Pasadena, CA 91109}

\altaffiltext{9}{Kavli Institute for Particle Astrophysics and
 Cosmology; Physics Department, Stanford University, and SLAC
 National Accelerator Laboratory, Stanford CA 94305}

\begin{abstract}

  Using observations in the COSMOS field, we report an intriguing
  correlation between the star formation activity of massive
  ($\sim10^{11.4}\,\msol$) central galaxies, their stellar masses, and
  the large-scale ($\sim$10 Mpc) environments of their group-mass
  ($\sim 10^{13.6}\,\msol$) dark matter halos.  Probing the redshift
  range $z=[0.2,1.0]$, our measurements come from two independent
  sources: an X-ray detected group catalog and constraints on the
  stellar-to-halo mass relation derived from a combination of
  clustering and weak lensing statistics.  At $z=1$, we find that the
  stellar mass in star-forming centrals is a factor of two less than
  in passive centrals at the same halo mass.  This implies that the
  presence or lack of star formation in group-scale centrals cannot be
  a stochastic process.  By $z=0$, the offset reverses, probably as a
  result of the different growth rates of these objects.  A similar
  but weaker trend is observed when dividing the sample by morphology
  rather than star formation. Remarkably, we find that
  star-forming centrals at $z\sim1$ live in groups that are
  significantly more clustered on 10 Mpc scales than similar mass
  groups hosting passive centrals.  We discuss this signal in the
  context of halo assembly and recent simulations, suggesting that
  star-forming centrals prefer halos with higher angular momentum
  and/or formation histories with more recent growth; such halos are
  known to evolve in denser large-scale environments.  If confirmed,
  this would be evidence of an early established link between the
  assembly history of halos on large scales and the future properties
  of the galaxies that form inside them.

\end{abstract}


\section{Introduction}

Understanding the form and evolution of the relationship between
galaxy stellar mass, galaxy color, and dark matter halo mass has
become a critical topic in galaxy formation. In
\cite{leauthaud_etal:12_shmr} (hereafter L12) we combined measurements
of the galaxy stellar mass function (SMF), galaxy clustering, and
galaxy-galaxy lensing in the COSMOS survey
(\citealt{scoville_etal:07}) to place constraints on the
stellar-to-halo mass relation (SHMR) at $0.2\le z\le 1.0$ using a halo
occupation analysis (HOD). In this
paper, we focus on the SHMR for massive galaxies, $\mgal\approx
10^{11-11.5}\,\msol$, within group-scale halos, $\mhalo\approx
10^{13.5}\,\msol$, across this same redshift range. Updating the L12
results, we now separately constrain the SHMR's for galaxies that are
actively star-forming and those that are passively evolving. We
compare these results with a sample of central galaxies identified in
an X-ray selected COSMOS group catalog (\citealt{george_etal:11}).

We define a dark matter halo with as having an overdensity 200 times
the mean cosmic density. All calculations assume a flat \lcdm\
cosmology of ($\om$,$\s8$,$\omb$,$n_s$,$h_0) =
(0.272,0.807,0.0438,0.963,0.72)$.

\section{Data}

The COSMOS sample that we use for clustering, lensing, and SMFs has
already been described in detail in L12. HOD anlaysis is performed on these
measurements. The main difference with
respect to L12 is that we now divide the sample into star-forming (SF)
and passive subsamples using the UVJ color-color cuts of
\cite{bundy_etal:10}.  
We use the same stellar masses as L12. These have been estimated using
the Bayesian code of \cite{bundy_etal:06} using a \cite{chabrier:03}
initial mass function. In our redshift range, there are 12,573 passive
and 41,682 SF galaxies in the COSMOS sample above our completeness
limits.

\begin{figure}
\epsscale{1.2} 
\vspace{2cm}
\plotone{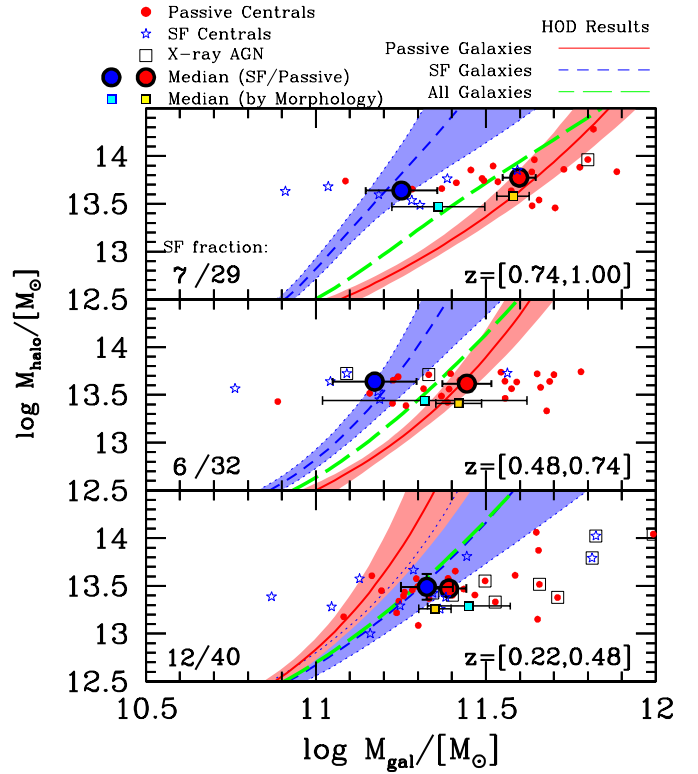}
\caption{ \label{shmr_groups} Evolution of the stellar-to-halo mass
  relation for group-scale halos. In each panel, the blue and red
  curves indicate the mean stellar mass as a function of halo mass for
  SF and passive central galaxies, respectively, from the HOD
  analysis. The shaded region around each curve is the 68\% confidence
  region on this mean. The green dashed curve shows the result for
  stellar-mass selected samples (no color cut) from L12. In each
  panel, all plot symbols represent results from the 
  group catalog. Small plot symbols show the
  halo and central galaxy masses individual groups; blue stars
  represent star-forming central galaxies, while red circles represent
  passive central galaxies. Objects with X-ray AGN activity are
  indicated with a black box. The larger points with error bars show
  the median mass of the central galaxies in the group catalog, and
  the uncertainty on that quantity. The green and yellow squares show
  the median values when splitting the sample by disk-like (green) and
  bulge-dominated (yellow) morphologies. We note that these results
  are obtained independently from the HOD results. }
\vspace{-0.5cm}
\end{figure}

We also use a COSMOS X-ray selected group catalog to select and study
central galaxies. Details regarding this group catalog can be found in
\cite{finoguenov_etal:07} and \cite{george_etal:11}\footnote{This
  group catalog is publicly available and can be found at
  http://irsa.ipac.caltech.edu/data/COSMOS/tables/groups/}. Halo
masses for these groups were determined in \cite{leauthaud_etal:10} by
calibrating the $L_X$-$\mhalo$ relation from weak lensing. To ensure a
clean sample of groups and centrals, we exclude potentially merging
systems, and groups near masked regions or with very few members
(\textsc{flag\_include}=1 in \citealt{george_etal:11}).  This sample
contains 129 groups out of 211 extended X-ray detections. We further
remove 18 groups with ambiguous identification of a central galaxy,
i.e., when the most massive group galaxy within the NFW-scale radius
(\citealt{nfw:97}) of the halo is not the most massive galaxy within
the virial radius (\textsc{mmgg\_scale\_mstar $\not=$
  mmgg\_r200\_mstar}).  At fixed redshift, the group catalog
constitutes a roughly halo-mass limited sample of dark matter
halos. We divide the data into three redshift bins that span
$z=[0.2,1.0]$. The specific redshift bins are the same as in L12 and
are shown in Fig.~\ref{shmr_groups}.  The mean logarithmic halo mass
in each redshift bin is 13.47, 13.59, and 13.75. We note that the mass
calibration of \cite{leauthaud_etal:10} assumes a halo mass definition
of 200 times the critical density. We have converted these values to
our fiducial halo definition by assuming the NFW density profile with
a concentration-mass relation given by
\cite{munoz_cuartas_etal:11}. We then rescale the masses from the
200-critical definition to the 200-mean (e.g., \citealt{hu_kravtsov:03}).

The central galaxies in our sample are well above the completeness
limit for COSMOS (Fig. 1 in L12), even for passive galaxies.  We also
check for AGN contamination, which we will discuss subsequently.

\section{Halo Occupation Analysis}

In \cite{leauthaud_etal:11a}, we presented a theoretical framework for
modeling combined measurements of the SMF, galaxy clustering, and
galaxy-galaxy lensing. This method is a more generalized version of
the traditional Halo Occupation Distribution (see, e.g.,
\citealt{cooray_sheth:02} for a review). Our HOD method utilizes these
three statistical measures to infer the number of galaxies within
halos as both a function of halo and galaxy mass.  In L12, we
implemented this formalism on stellar-mass defined samples within
COSMOS. Our analysis constrains the halo occupation of both central
and satellite galaxies, but in this paper we focus exclusively on
central galaxies within group-scale halos.
We constrain a SHMR for both passive and SF central galaxies such that
the total number of central galaxies per halo is unity. This result
is obtained independent of the SHMR constrained from the group
catalog. In L12 we assumed that every halo has one central galaxy;
here we require that the sum of mean occupation of passive and SF
central galaxies is unity. In a companion paper (J. Tinker et al., in
preparation), we present full details of our measurements and our
model fits. Our results focus on the relative clustering of groups at
$\sim 10$ Mpc. Due to the small area of COSMOS, the integral
constraint can affect the clustering of objects at our scale of
interest (L12). However, it will not alter the relative clustering of
two samples in the same volume (L12), which is the quantity of
interest here.

\section{Results}

\subsection{Stellar-to-Halo Mass Ratios}

Fig.~\ref{shmr_groups} shows the constraints on the SHMR for passive
and SF galaxies for each redshift bin. Results from L12 for the full
stellar-mass limited samples are shown for comparison. At $z=0.9$,
there is a clear difference between the stellar masses of SF and
passive central galaxies in groups of similar halo mass. At
$\mhalo=10^{13.7} \msol$, the difference is 0.4 dex. This is
qualitatively consistent with the trends seen in AEGIS groups at lower
halo mass (\citealt{woo_etal:12}). At lower redshift, however, this
difference gradually goes away. In the lowest redshift bin, the SHMR
for SF and passive galaxies cross at fixed halo mass.

This evolution is confirmed in the galaxy group sample: At high
redshift, there is a 3-$\sigma$ difference in the median central
galaxy mass between passive and SF centrals. Errors for this quantity
are calculated by bootstrap resampling of the stellar masses within
each subsample. The median galaxy masses are also in good agreement
with those derived from the halo occupation analysis. At lower
redshifts, the difference in the passive and SF galaxy masses gets
monotonically smaller. The median masses do not cross over, as they do
in the SHMRs; there is a discrepancy between the results for the
passive subsample in the lowest redshift bin. However, results from SDSS
demonstrate that this cross over has indeed occurred by $z=0$
(\citealt{mandelbaum_etal:06_gals, more_etal:11}). The results from
the group catalog are qualitatively similar if one breaks the catalog
up by morphology\footnote{The `spheroidal' classification of \S 3.4.1
  in \citealt{bundy_etal:10}.} (as shown in Figure
\ref{shmr_groups}). The galaxies with X-ray AGN activity, either in
the XMM or Chandra observations are indicated on the plot. The low
number of such objects, and the (lack of) correlation with star
formation estimates indicates that AGN contamination is not playing
any role in the observations. Removing these objects from the sample
does not shift the medians beyond their 1-$\sigma$ errors.

At $z=[0.22,0.48]$, the discrepancy in the SHMR values and those
obtained from the groups for passive galaxies is a 2.4-$\sigma$
difference based upon creating Monte Carlo samples of halos using all
elements in the MCMC chain but with the same mass distribution as the
groups sample.
The large-scale clustering amplitude of all structure in the low-$z$
bin is below average (L12), which could drive systematic errors
in the HOD results. It is also possible that the halo mass function
assumed in the HOD analysis (\citealt{tinker_etal:08_mf}) is not the
same as the true mass function in that patch of sky, also resulting in
systematic biases. However, we note that the groups catalog is a
subset of all X-ray groups within COSMOS, while the SHMRs are derived
from statistics on the full sample of galaxies. Also, $z=0$
measurements of the SHMR using lensing
(\citealt{mandelbaum_etal:06_gals}) and satellite kinematics
(\citealt{more_etal:11}) find that the halo masses for massive red
galaxies are higher than those of massive SF central galaxies,
following the evolutionary trend seen in the HOD results.

\subsection{Clustering by Central Galaxy Type}

\begin{figure}
\epsscale{1.2} 
\plotone{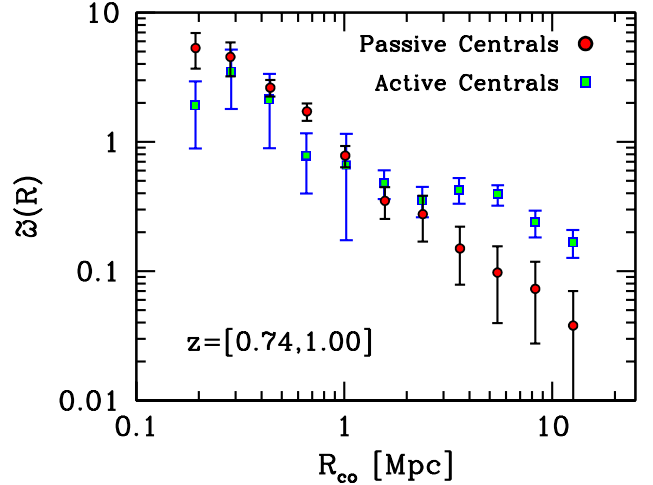}
\vspace{-2.cm}
\caption{ \label{wtheta_groups} The cross-correlation function of the
  X-ray groups with all galaxies in the defined redshift range. The
  $x$-axis is the comoving projected separation between pairs. The
  $y$-axis, $\omg$, has an arbitrary normalization, thus the relative
  amplitude is the key quantity (see text for details). Black/red
  circles represent groups with passive central galaxies; blue/green
  squares represent groups with star-forming central galaxies. Note
  that the groups with passive centrals are slightly more massive than
  the groups with star forming centrals. }
\end{figure}

Fig.~\ref{wtheta_groups} shows the cross-correlation between groups
and all galaxies. We split the groups into samples with SF and passive
centrals, and cross-correlate each set of centrals with all
galaxies. Both samples are in the z=[0.74, 1.00] redshift bin, with a
magnitude cut of $i_{F814W}=24$. Galaxies brighter than this threshold
have reliable photometric redshifts with errors $\sim 0.03$ (see
figure 2 in \citealt{george_etal:11}). Because most of the redshifts
of the central group galaxies are known with spectroscopic precision
(most are sampled within the zCOSMOS survey), we can measure the
real-space projected clustering, which has higher signal-to-noise
relative to a simple angular cross-correlation. We denote this
clustering statistic $\omg$. Details are given in
\cite{padmanabhan_etal:09}. Briefly, we calculate the projected
comoving separation between each group-galaxy pair from the angular
separation and the redshift of the group. We restrict all pairs to lie
within a redshift interval of $\pm 0.09$, or 3-$\sigma$ of the photo-z
error. To properly normalize $\omg$ requires detailed information of
the photo-z error distribution function, but since we are only
concerned with the relative clustering between two spectroscopic
samples cross-correlated with the same photometric sample, this step
is unnecessary. We measure the angular cross-correlation, $\wth$, for
the same samples as a cross-check on our results. Errors are obtained
by bootstrap resampling of the groups and recalculating $\omg$ or
$w(\theta)$ for each bootstrap sample.

A scale of importance is the 1 Mpc scale (comoving), roughly the
virial radius of the groups. Inside this scale, the cross-correlation
probes the number of satellite galaxies within the groups. Outside
this scale, the cross-correlation probes the large-scale bias of the
groups, which is an indicator of their environment. In the
measurements of Fig.~\ref{wtheta_groups}, this scale marks the
bifurcation in the two correlation functions. Outside this scale, the
passive-central groups have a lower large-scale bias, indicating that
these halos have formed in lower-density environments. Inside this
scale, the correlation functions for the groups differ at the
$1\sigma$ level, with the SF-centered groups having lower clustering,
but the large errors prevent meaningful interpretation.

\begin{figure}
\epsscale{1.2} 
\vspace{-3cm}
\plotone{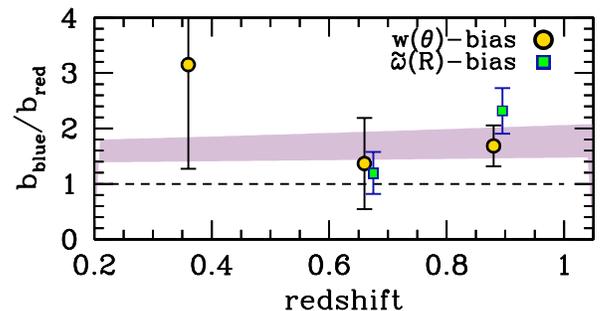}
\vspace{-1.cm}
\caption{ \label{relative_bias} The relative large-scale bias of SF
  and passive-centered groups ($R_{\rm co}>1$ Mpc). Circles and
  squares represent bias obtained from the ratio of the angular
  cross-correlation function, $\wth$, and the real-space
  cross-correlation function, $\omg$, respectively. Valid bias values
  cannot be obtained for the $z=0.36$ $\omg$ measurements. The shaded
  band is the predicted range for relative bias of halos separated by
  angular momentum, with high angular momentum halos being more
  clustered (see the discussion in \S 5). The dashed line indicates no
  difference in the clustering between the two subsamples. }
\end{figure}

For each redshift bin, we calculate the bias relative to an
(arbitrarily normalized) non-linear matter correlation function
calculated using the \cite{smith_etal:03} fitting function. We
calculate bias using bins at $R_{\rm co}>1$ Mpc or $\theta>80$
arcsec. We estimate the covariance matrix by bootstrap resampling of
the groups with replacement. We use 200 bootstrap samples. Due to the
low number of groups, the clustering signal around each group can be
considered independent. We use the full covariance matrix to obtain
the bias and its error. The relative bias of SF-centered and
passive-centered groups is shown as a function of redshift in
Fig.~\ref{relative_bias}. We show bias measurements from both $\omg$
and $\wth$. For the former, a bias measurement is not possible at
$z=0.36$ due to noise in the $\omg$ measurements for both
subsamples. At $z=0.88$, both measures indicate that the SF-centered
groups have significantly enhanced clustering. At $z=0.66$, the
relative bias is above unity, but this detection is not significant
given the errors. At $z=0.36$, the angular clustering yields a
1-$\sigma$ detection of elevated clustering in the SF-centered
groups.

\section{Discussion}

Proposed quenching mechanisms for massive galaxies, such as major
mergers (i.e., \citealt{hopkins_etal:08b}) or AGN (i.e.,
\citealt{de_lucia_blaizot:07}) essentially remove galaxies from the
star-forming sequence and place them in the red sequence. The
efficiency of such processes is correlated with halo mass. Some
research suggests that cooling flows may lead to episodic star
formation in central galaxies (\citealt{liu_etal:12}).

The substantial difference in the SHMRs for SF and passive galaxies at
the group-scale halo masses at $z=1$ has many implications. It implies
that star formation is not stochastic in these objects: if massive
central galaxies underwent periodic episodes of star formation
followed by longer-term quiescence, the galaxies at fixed halo mass
would have the same mass regardless of color. The results also imply
that massive quenched galaxies had far different growth histories than
those that are forming stars at $z=1$. A scenario in which galaxies at
fixed halo mass grow on a common star-forming sequence, with a
quenching mechanism that removes these galaxies from this sequence,
would make passive central galaxies {\it less} massive than SF central
galaxies. This is the opposite of what is observed at $z=1$. To be
consistent with our observations, passive central galaxies at $z=1$ form
their stars rapidly at high redshift, essentially getting `ahead of
the growth curve' relative to central galaxies that are still forming
stars by $z=1$. At high redshift, central galaxies essentially
``knew'' they would be quenched by $z=1$.

Fig.~\ref{wtheta_groups} indicates that color-selected groups
represent special subsets of objects at this halo mass scale. The
current growth rate (indicated by galaxy color) and growth history
(probed by total stellar mass) of the central galaxy is correlated
with large-scale environment. A similar effect is seen in dark matter
halos in N-body simulations, an effect called assembly bias. For
massive systems, younger halos exist in more dense environments
(\citealt{wechsler_etal:06, dalal_etal:08}). The environment (and
formation history) of massive halos is also correlated with angular
momentum of dark matter halos (\citealt{wechsler:01, bett_etal:07,
  gao_white:07}) such that high-spin halos are more clustered than
low-spin halos. This effect goes away below $\sim 10^{12} \msol$. The
shaded region in Fig.~ \ref{relative_bias} is the numerical result
from \cite{bett_etal:07}. The lower limit is the bias of the top 20\%
of halos, ranked by angular momentum, relative to all halos. The upper
limit is the bias of the top 20\% of halos relative to the lowest 20\%
of halos (see their Fig.~20\footnote{We assume that assembly bias is
  fixed for halos with the same $\sigma(M,z)$, thus we convert their
  $z=0$ results, which are plotted as a function of $M$, to
  $\sigma(M,z)$ and interpolate the assembly bias at the values of
  $\sigma(M,z)$ for the groups samples at each redshift.}). The proper
comparison to our measurements will lie in between.

Recent hydrodynamic simulations of galaxy formation indicate that
galaxy morphology is correlated with the angular momentum gained from
the larger-scale environment around the halo at early epochs
(\citealt{sales_etal:11}). If the angular momentum of dark matter and
baryons are connected, massive halos with central disk galaxies should
have enhanced clustering, in agreement with the results in Fig.~
\ref{wtheta_groups}.
\cite{sales_etal:11}
also find that the disk galaxy masses are lower than their spheroidal
counterparts. There are, however, substantive differences between the
Sales simulations and our results: they show results for less-massive
halos at $z=0$, rather than group-scale systems at $z=1$. They
conclude that there is little correlation between morphology and $z=0$
halo spin, but there is little correlation between halo spin at early
and late epochs for their halo masses
(\citealt{vitvitska_etal:02}). However, the existence of assembly bias
implies that more massive halos retain memory of the angular momentum
at the epoch of galaxy formation.  Further investigation is required
at higher masses and redshifts.

However, one need not invoke angular momentum to achieve both the
relative clustering and relative masses of passive and SF
centrals. \cite{conroy_wechsler:09} demonstrate that stellar mass
growth peaks at a halo mass of $\sim 10^{12} \msol$, weakly dependent
on redshift, but the star formation efficiency at that peak decreases
with cosmic time. In this scenario, central galaxies within
late-forming halos would lag behind those in early-forming halos, and
have enhanced clustering. This toy model does not explain the
morphology dependence, or the difference in instantaneous SF rates at
$z=1$, but does provide a connection between halo formation history
and galaxy properties.

There have been many attempts to find assembly bias in the $z=0$
galaxy distribution. \cite{tinker_etal:08_voids} and
\cite{tinker_etal:11_groups} find no evidence for assembly bias for
galaxies below the knee in the SMF or luminosity
function. The assembly biases in the low mass and high mass halo
populations are driven by disparate physical mechanisms. Younger halos
form in denser environments at high mass through the statistics of
Gaussian random fields. At low mass, {\it older} halos form in denser
environments due to tidal forces and interactions with nearby massive
objects (\citealt{dalal_etal:08}). It is plausible that these two
mechanisms may have different levels of impact on galaxy formation.

\cite{wang_etal:08_assembly_bias} find that $z=0$ group-mass halos in
SDSS with redder total galaxy content (centrals and satellites
combined) are more clustered than groups with bluer galaxies. It is
not clear how the clustering signal in $z=1$ COSMOS data could reverse
if the most-clustered halos at one redshift remain the most clustered
at a lower redshift. The \cite{wang_etal:08_assembly_bias} detection
is mitigated by the lack of independent constraints on the halo mass;
in their group-finding algorithm, the halo mass is estimated
statistically by assuming a 1:1 correspondence between total group
stellar mass and halo mass, with no scatter. \cite{berlind_etal:06},
using a different group-finding algorithm (but the same data set),
find the opposite signal: groups with bluer central galaxies are the
ones that are more clustered. Both these methods rely on inferring
halo mass statistically from the galaxies within them; our X-ray
detections and lensing masses are more legitimate for detecting
assembly bias.

From $z=1$ to $z=0$, the SHMRs evolve quite differently depending on
star formation activity. By $z=0.36$, the mean relations have crossed
and passive central galaxies live in higher mass halos than SF central
galaxies at fixed mass. This inversion is also consistent with results
from $z=0$ studies (\citealt{mandelbaum_etal:06_gals,
  more_etal:11}). Star forming galaxies grow by a factor of $\sim$2
using the star formation rates of \cite{noeske_etal:07a} from $z=0.88$
to $z=0.36$. Group-mass halos also grow by a factor of $\sim$2, thus
central galaxies grow as fast as their host halos. For quenched
galaxies, their growth rates are slower than that of their host halos,
plausibly causing the inversion of the SHMR seen in
Fig.~\ref{shmr_groups}.

At $z\gtrsim 1$, our results imply that the process that shuts down
star formation in massive galaxies cannot be explained by a stochastic
process that is a function of halo mass. Rather, the interplay between
the dark matter halo and the surrounding environment, including the
tidal fielde, strongly influences the
fate of the galaxy forming within it.

\acknowledgements

We thank Charlie Conroy and Tom Theuns for useful discussions. This
work was supported by World Premier International Research Center
Initiative (WPI Initiative), MEXT, Japan.  The initial HST-COSMOS
Treasury program was supported through NASA grant HST-GO-09822.  We
gratefully acknowledge contributions of the entire COSMOS
collaboration consisting of more than 140 scientists.  More
information on COSMOS is available at {\bf
  \url{http://cosmos.astro.caltech.edu/}} and the data archive is at
IPAC/IRSA. JR was supported by JPL, run under contract for
NASA by Caltech.


\break

\end{document}